\documentclass{pazh_engl}
\usepackage{epsf}
\usepackage{flushrt}
\usepackage{graphicx}
\usepackage{natbib}

\begin{document}
     
\sloppypar

\title{Constraints on the luminosity of the stellar remnant 
in SNR1987A}

\author{P.Shtykovskiy\inst{1,2}, A.Lutovinov \inst{1,2}, M.Gilfanov\inst{2,1},
R.Sunyaev\inst{1,2}}

\offprints{pav\_sht@hea.iki.rssi.ru}

\institute{Space Research Institute, Russian Academy of Sciences,
           Profsoyuznaya 84/32, 117997 Moscow, Russia
\and
           Max-Planck-Institute f\"ur Astrophysik,
           Karl-Schwarzschild-Str. 1, D-85740 Garching bei M\"unchen,
           Germany
}
\date{Received 18 November 2004/Accepted}

\authorrunning{Shtykovskiy et al.}
\titlerunning{Constraints on the luminosity of the stellar remnant 
in SNR1987A}

  \abstract{We obtain photometric constraints on the luminosity of the stellar
remnant  in SNR1987A using XMM-Newton and INTEGRAL data.
The upper limit in the 2--10 keV band based on the
XMM-Newton data is
$L_X\la5\times 10^{34}$erg/s. We note, however, that the optical depth
of the envelope is still high in the XMM-Newton band, therefore, this
upper limit does not constrain the true unabsorbed luminosity of the
central source. The optical depth is expected to be small in the hard
X-ray band of the IBIS telescope aboard the INTEGRAL
observatory, therefore it provides an unobscured  look at the stellar
remnant. We did not detect statistically significant emission
from  SN1987A in the 20-60 keV band with the upper limit of
$L_X\la1.1\times 10^{36}$erg/s.
We also obtained an upper limit on the mass of radioactive $^{44}$Ti
M($^{44}$Ti)$\la10^{-3}$M$_{\sun}$.
   }

   \maketitle

%

\section{Introduction.}
\label{sec:intro}
Owing to its proximity, Supernova SN1987A in the Large
Magellanic Cloud provides a unique possibility to study the physics
of a supernovae and supernova remnants.

Today it reveals itself as an extended object experiencing rapid
evolution. The observed X-ray emission of the supernova is dominated by the
interaction of the shock wave with the matter produced by the wind
from a supergiant phase of its progenitor.  As a result of this
interaction the supernova was gradually increasing its luminosity
during the past decade. This was followed by a strong brightening of
the supernova on December 2002 when the shock wave encountered the
so-called inner ring (see \citet{chandra} and references therein).

Based on the Balmer lines in the optical spectrum, identification of a 
massive blue supergiant progenitor
\citep{progenitor} and the neutrino burst \citep{neutrinoburst} coincident
with the SN event, SN1987A was classified as a Type II core-collapse
supernova.
The mass of the progenitor was estimated to be
$\approx15-20$M$_{\sun}$ \citep{progenitormass}.
\citet{sunyaevhexe} detected strong very hard X-ray source after about
169 days from the explosion but this emission was connected with
comptonisation of gamma-ray lines of Co$^{56}$ decay in the optically
thick envelope. The lines themselves became visible later \citep[e.g.][]{gammalines}.

The collapse of such a massive star should lead to the formation of a
neutron star or a black hole. If a neutron star was born with strong
magnetic field and was rapidly rotating, it should appear as a luminous
Crab-like X-ray source or as an accreting pulsar, which usually also
have very hard X-ray spectrum with maximal contribution to
luminosity in the INTEGRAL band.
Neutron star without strong magnetic field or a black hole might
become a bright X-ray source due to the accretion of the matter 
of the dense envelope.
\citet{chandra} obtained from an imaging analysis of the Chandra data 
an upper limit L$_X$(2-10 keV)$<1.5\times10^{34}$ erg/s on the
luminosity of the central source.
However the envelope continues to be optically thick and there is 
a possibility that its optical depth  is still high in
the standart X-ray band and does not permit us to
observe the central source directly.
It is important to check that it is not specially bright in the hard
X-ray band, where the envelope is transparent both for the Thomson
scattering and photoabsorption.
In addition we decided to check the available data obtained by the
XMM-Newton observatory, which has higher sensitivity for photons
with energies $\ga5$keV than Chandra.

Here we present results of analysis of INTEGRAL and XMM-Newton observations of
SNR1987A and constrain the luminosity of the stellar remnant embedded
in the envelope based on its X-ray spectrum.

\section{Data reduction and results}
\label{sec:data}

\subsection{XMM-Newton}

SNR1987A was observed by XMM-Newton in September and November 2000,
April 2001 and May 2003. These observations were processed with the
Science Analysis System (SAS) v6.0.0. The data were screened for the
soft proton flares by  removing the time intervals where the count
rate above 10 keV significantly exceeded the mean level.  For the
pointing in April 2001 because of numerous soft proton flares 
the background threshold was chosen higher than the
quiescent background typical for time periods without flaring
activity. To be sure that this does not influences our results, we
repeated analysis with different thresholds on background level and a
background subtraction from different regions on detector and found no
significant variations in the obtained results.

\begin{figure*}
\hbox{
\includegraphics[width=0.5\textwidth]{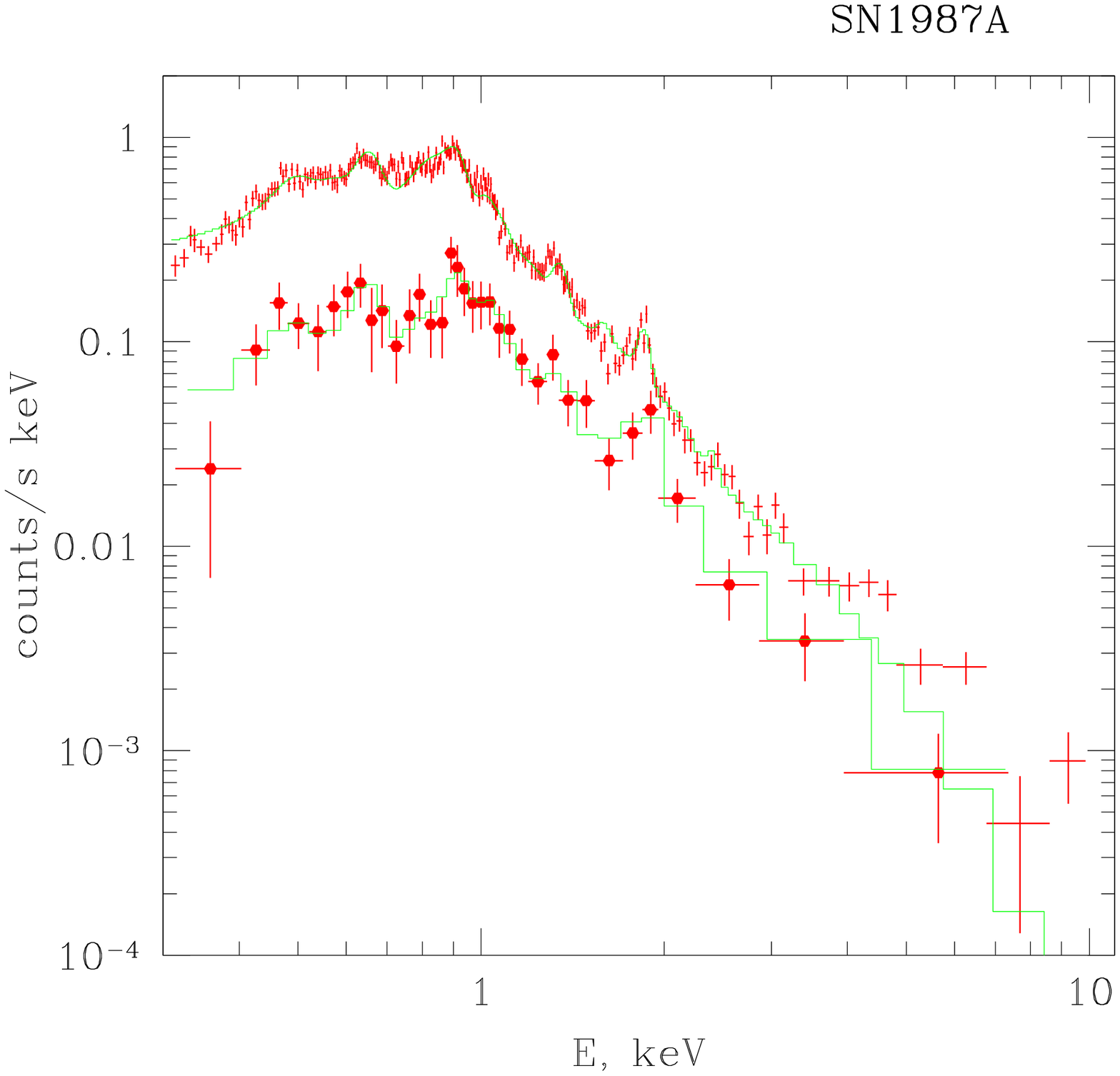}
\includegraphics[width=0.5\textwidth]{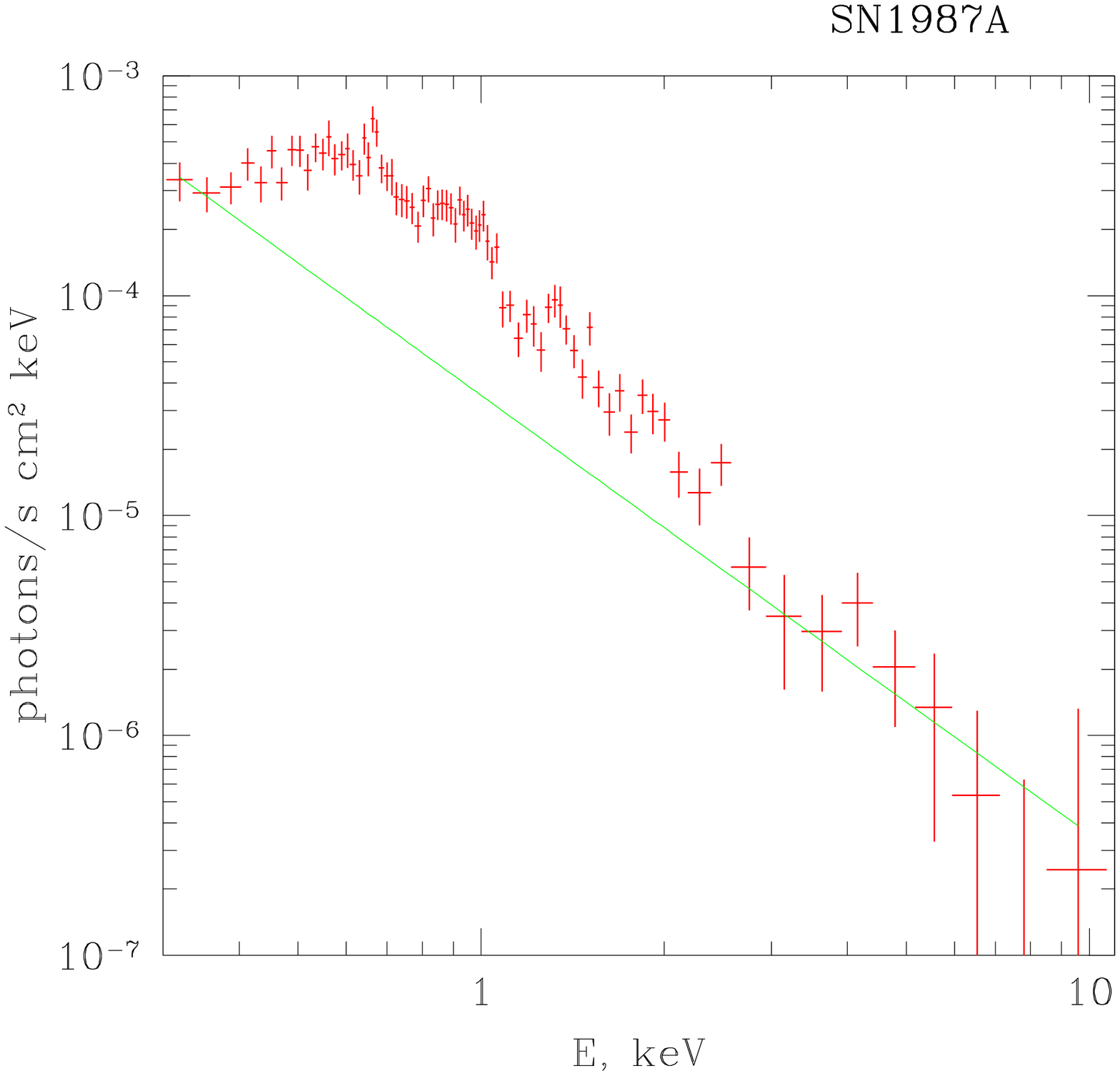}
}
\caption{{\em Left:}
EPIC PN spectra of the SNR1987A. The upper curve corresponds
to pointing on May 2003, the lower to pointing on September 2000.
{\em Right:}
Deconvolved photon spectrum of SNR1987A on April 2001 with the powerlaw model
describing the data on energies above 4keV.
}  
\label{fig:xmmspectra}
\end{figure*}  

The angular resolution of the XMM-Newton, unlike that of the Chandra, is
insufficient to resolve the
remnant, and it appears as a point-like source on the extracted images. 
However, XMM-Newton is more sensitive to the
photons with energy $>5$ keV and therefore allows to reconstruct the
sources spectra to higher energies.
 The EPIC spectra were extracted from a $30\arcsec-40\arcsec$ circle
regions around the SNR. Corresponding background spectra were
extracted from the nearby regions.

The spectra of the SN1987A remnant obtained with XMM-Newton on September
2000 and May 2003 together with NEI model fit (the plane-parallel
shock model \citep{borkowski}
where the plasma has not reached the collisional ionization equilibrium)
are shown in ~Fig.\ref{fig:xmmspectra}. As is
evident from the figure, the luminosity of the remnant increased by
a factor of $\approx$3 during three years. This is expected from
the interaction of the supernova shock with the matter left from the
stellar wind of a supergiant progenitor.
It seems that there is some excess of hard X-ray emission in
comparison with NEI model for pointing on May 2003.
Such a high energy tail could be associated with either a central source 
or a synchrotron radiation from the shell of the remnant \citep[e.g.][]{synchrotronrem}.
We focus on the possible association of the high energy excess
with the central source.
Detailed analysis of the supernova remnant spectra is not a subject
of this paper and could be found in e.g. \citet{chandra,aschenbach}.

To obtain a conservative upper limit on the central source luminosity
we shall assume that the total flux observed at the high
energy part of the remnant's spectrum is produced by the central
source. For this analysis we
used the data of the early pointings (2000 and 2001 years), 
as a contribution of the
emission originating from the interaction of the shock with the medium
would be smaller for them.  Assuming the spectrum of the central source to be
a powerlaw with a photon index 2.0, we obtained the upper limit
(1$\sigma$) for its 2-10 keV luminosity L$\la5\times 10^{34}$erg/s,
assuming a distance to the source $d=50$ kpc. The obtained value is
in the same range, but somewhat higher than the upper limit obtained by Chandra
from the imaging analysis \citep{chandra}.

\subsection{INTEGRAL}

The international gamma-ray laboratory INTEGRAL \citep{wink03}
observed the Large Magellanic Cloud several times on January 2003 with a
total exposure about 1 Mln sec. In our analysis we used publicly
available data obtained by the ISGRI detector of the IBIS telescope of
the observatory. ISGRI is sensitive to the photons in the E$>20$ keV
energy band. The data of INTEGRAL/IBIS/ISGRI were processed with the
method developed by Eugene Churazov and described in 
Revnivtsev et al. (2004). Detailed analysis of
Crab nebula observations suggests that with the approach and software
employed, the conservative estimation of the uncertainty in measurements
of absolute fluxes from the sources is about 10\%.

\begin{figure*}
\begin{center}
\includegraphics[width=0.8\textwidth,clip=true]{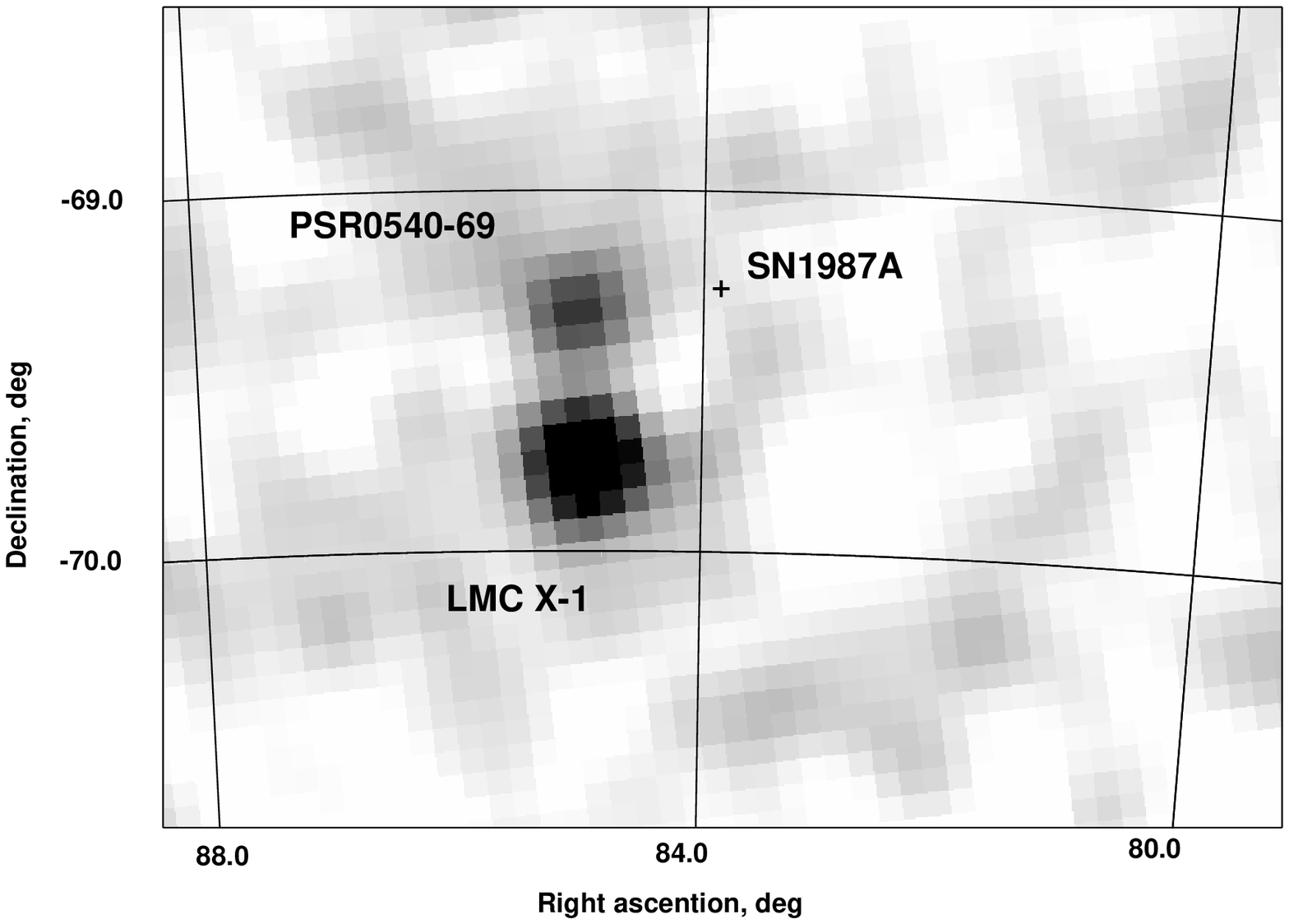}
\end{center}
\caption{Map of the part of LMC around of the SN1987A position,
obtained with INTEGRAL in the 20-60 keV energy band. Position of
SN1987A is shown by the cross.} \label{ima}
\end{figure*}

To investigate the hard X-ray emission from the SN1987A remnant we
analyzed data obtained with INTEGRAL/IBIS/ISGRI in the 20-60 keV
energy band, where the sensitivity of the ISGRI detector is
maximal. We reconstructed the image of the central part of the Large
Magellanic Cloud in this energy band (see Fig. \ref{ima}) and did not
find statistically significant emission from the remnant with the
upper limit of
$F_X\la3.7\times10^{-12}$ erg/s/cm$^2$ (2$\sigma$, assuming
a Crab-like spectrum), that corresponds to the luminosity of
$L_X\la1.1\times 10^{36}$erg/s.

\begin{figure*}
\begin{center}
\includegraphics[width=0.8\textwidth,clip=true]{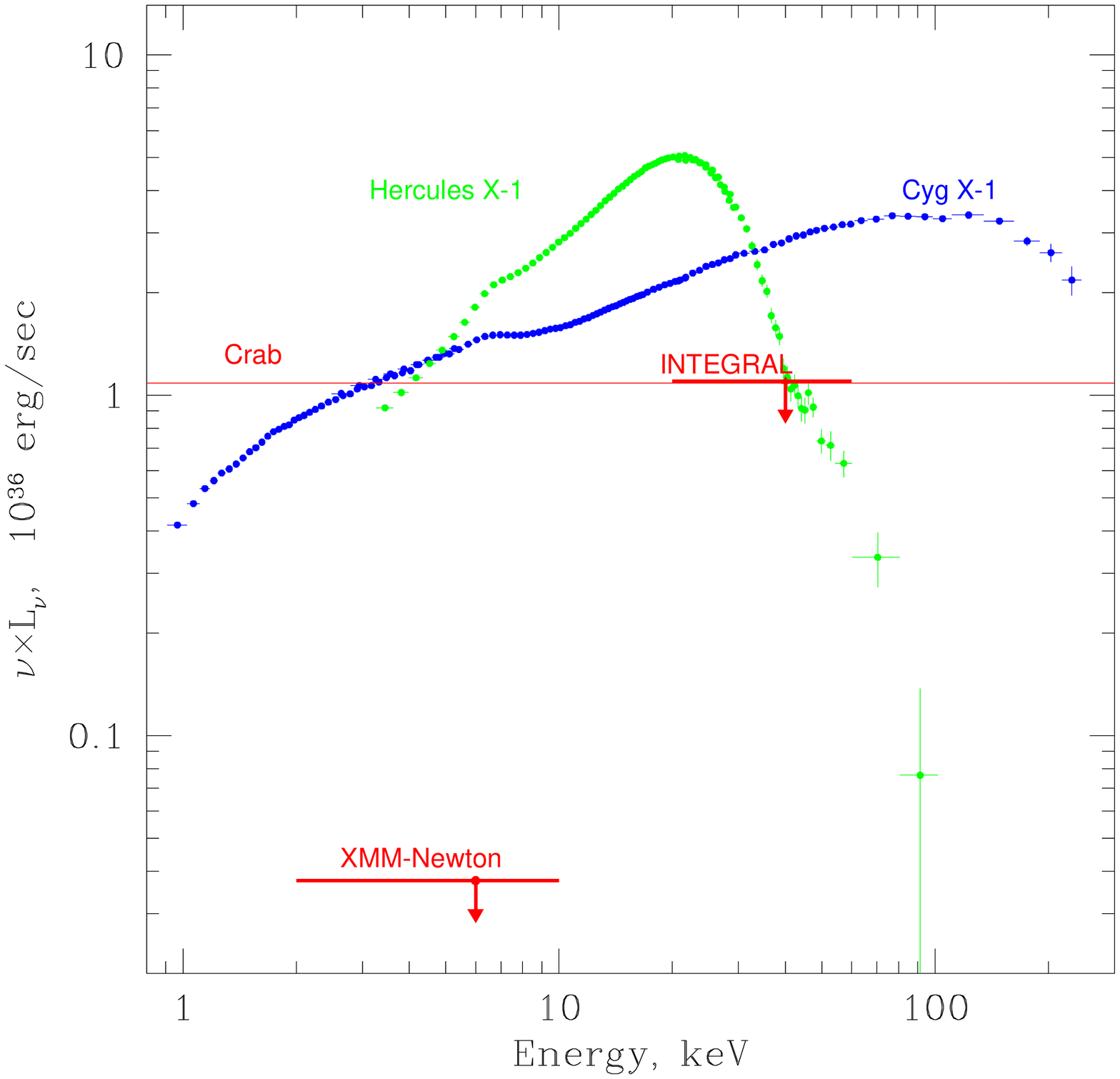}
\end{center}
\caption{The typical spectra produced by 
accreting X-ray pulsar, accreting black hole \citep{cygx1} and 
rotation-powered Crab-like pulsar.
All spectra are normalised to the luminosity of $2.5\cdot10^{36}$erg/s in the
1-12 keV energy band.
This corresponds to the bolometric optical and infrared luminosity
\citep{bolom}, which could be interpreted as an upper limit on the 
central source luminosity absorbed by the envelope.
Upper limits on the luminosity of central source 
in SNR1987A obtained by the XMM-Newton and INTEGRAL observatories
are also shown.
} \label{fig:limit}
\end{figure*}

To compare this result with results of Chandra and XMM-Newton
observatories we can extrapolate it to the 2-10 keV energy band.
The spectra of pulsars in the supernova remnants are usually
approximated by a powerlaw model with a slope of $\sim1.5-2$. 
For such values of the photon index we can estimate
the central source luminosity in the 2-10 keV energy band corresponding to the
INTEGRAL upper limit as $L_X\la(0.6-1.6)\times 10^{36}$erg/s.

\subsection{Optical depth of the envelope}
The envelope of the supernova at early times 
was optically thick both for photoabsorption and Thomson scattering.
This, for example, follows from the observations of Rentgen observatory
on the KVANT module of MIR station
 \citep[][see also \citet{snexpectedem}]{sunyaevhexe}.
As the envelope expands and its density decreases, its
optical depth decreases with time as $\tau\propto t^{-2}$. 
Given the \citet{snlateemission} model for the photoabsorption optical
depth of the envelope, and assuming t$^{-2}$ behaviour, we obtain 
$\tau_{photo}\sim7$ for a 5 keV photons at the epoch of XMM-Newton obervations.
The photoabsorption optical depth reaches unity only on energies  
$\sim$11-15 keV.
Thus, the optical depth is still high in the energy bands of Chandra
and XMM-Newton.
Therefore the upper limits provided by these observatories 
correspond to the small fraction of the central source emission 
which escaped the envelope,
while the real central source luminosity could be much
higher.

Similar estimate of the optical depth for 40 keV photons give value 
$\tau_{photo}\sim0.02$, therefore the hard X-ray radiation escapes
the envelope without absorption.
For the Thomson optical depth we obtained also very low values, 
$\tau_T\sim0.005-0.1$ depending on the model of matter distribution 
in the envelope.
Was the $\tau_T$ close to unity or even a bit higher, it would not
affect the flux in 20-60 keV energy band because of small
energy change for one scattering, $\Delta E/E\approx h\nu/m_ec^2\sim0.1$
\citep{pozdnyak}.
Thus, the INTEGRAL observations offer us a direct look at the
central source and its upper limit constraints the real luminosity 
of the stellar remnant.

Absorbed emission in the standart X-ray band is reemitted in the
optical and infrared bands and an upper limit on the central source 
luminosity could be estimated based on a bolometric luminosity of the
SNR in these energy bands.
For example, \citet{bolom} estimated the bolometric luminosity of the remnant
log(L)$\sim36.1-36.4$ after the 3600 days from the explosion.
This value could be interpreted as an upper limit on the 
central source luminosity corrected for the absorption, what is close to
a limit obtained from the INTEGRAL data.
Obviously, the upcoming Spitzer data will provide an interesting
estimates on the luminosity of the central source, absorbed and
reradiated in the infrared band by the optically thick for the
standart X-rays envelope.

\subsection{Limit on the nucleosynthesis of $^{44}$Ti}
While the early (except for the first few weeks) bolometric 
 luminosity of the supernova was powered by 
the $^{56}$Co decay (t$_{1/2}\sim$77 days), at later times 
elements with longer decay times, like $^{44}$Ti (t$_{1/2}\sim$60 years), 
give the main contribution.
Decay of the $^{44}$Ti gives rise to
two gamma photons with energies 67.9 keV and 78.4 keV.
Simple estimations on the fluxes of these lines give value
$\sim3\cdot10^{-13}$erg/s/cm$^{2}$ \citep[see e.g.][]{titan},
assuming mass of titanium  M($^{44}$Ti)$=10^{-4}$M$_{\sun}$.
To estimate the mass of the synthesed $^{44}$Ti, we calculated
fluxes observed by INTEGRAL in the 61-73 keV and 73-88 keV energy bands,
F$_X$(61-73 keV)=$4.30\pm1.25\cdot10^{-12}$erg/s/cm$^{2}$  
and F$_X$(73-88 keV)=$2.50\pm1.60\cdot10^{-12}$erg/s/cm$^{2}$.
In the first case the flux
has a formal significance of 3.4$\sigma$, however we 
consider it insufficient to claim a reliable detection of the line emission.
Therefore to estimate the mass of $^{44}$Ti, we assume both fluxes to
be upper limits and obtain M($^{44}$Ti)$\la10^{-3}$M$_{\sun}$.

\section{Conclusion}
\label{sec:summary}
Based on the data of XMM-Newton and INTEGRAL, we provide broadband
spectroscopic constraints
on the luminosity of the central source in the SN1987A. 
Assuming that a hard energy tail of the XMM-Newton spectrum of the
remnant is dominated by a central source, we obtained an upper limit 
L$\la5\times 10^{34}$erg/s  for 2-10 keV central
source luminosity uncorrected for absorption,  
which is somewhat higher than the upper limit obtained by Chandra from 
imaging analysis.
INTEGRAL did not detect any statistically significant emission from 
the remnant and provided an
upper limit for 20-60 keV luminosity of $L_X\la1.1\times 10^{36}$erg/s,
what corresponds to 2-10 keV luminosity $L_X\la(0.6-1.6)\times
10^{36}$erg/s assuming Crab-pulsar like spectrum.
The obtained upper limits on the possible central source luminosity
in comparison with the typical spectra of Crab-like pulsar, 
accreting X-ray pulsar and accreting black hole are shown on Fig.~\ref{fig:limit}.

Based on the existing calculations of the photoabsorption optical
depth of the envelope, we show that absorption in the standart X-ray band 
of the XMM-Newton and Chandra is still high and obtained upper limit
corresponds to luminosity not corrected for absorption, which
could be only a small fraction of a real luminosity. 
The optical depth in the INTEGRAL energy band is significantly less
than unity and obtained upper limit constrain real luminosity of a 
stellar remnant.

We also obtained an upper limit on the mass of $^{44}$Ti  
 M($^{44}$Ti)$\la10^{-3}$M$_{\sun}$.

\begin{acknowledgements}
The authors thank Eugene Churazov for developing of methods of
analysis of the IBIS data and software. 
The authors also would like to thank the referee S.A. Grebenev
for helpful comments. This research has made use of
data obtained through the INTEGRAL Science Data Center (ISDC),
Versoix, Switzerland, Russian INTEGRAL Science Data Center (RSDC),
Moscow, Russia and XMM-Newton science archive. 
This work was supported by grants of Minpromnauka
NSH-2083.2003.2 and the program of Russian Academy of Sciences
``Non-stationary phenomena in astronomy''. AL acknowledge the support
of RFFI grant 04-02-17276.
\end{acknowledgements}

\end{document}